\documentclass[prb,
superscriptaddress,showpacs,amsmath,amssymb]{revtex4}

\begin{document}

\title{Thermodynamics of black and white holes in ensemble of Planckons}

\author{G.E.~Volovik}
\affiliation{Landau Institute for Theoretical Physics, acad. Semyonov av., 1a, 142432,
Chernogolovka, Russia}

\date{\today}

\begin{abstract}
The Tsallis-Cirto non-extensive statistics with $\delta=2$ describes the processes of splitting and merging of black holes and their thermodynamics.\cite{Volovik2025a,Volovik2025d} 
Here we consider a toy model, which matches this generalized statistics and extends it by providing the integer valued entropy of the black hole, $S_{\rm BH}(N)=N(N-1)/2$. In this model the black hole consists of  $N$ the so-called Planckons  -- objects with reduced Planck mass $m_{\rm P}=1/\sqrt{8\pi G}$ -- so that its mass is quantized, $M=Nm_{\rm P}$. The entropy of each Planckon is zero, but the entropy of black hole with $N$ Planckons is provided by the $N(N-1)/2$ degrees of freedom -- the correlations between the gravitationally attracted Planckons. This toy model can be extended to a charged Reissner-Nordstr\"om (RN) black hole, which consists of charged Planckons. Despite the charge, the statistical ensemble of Planckons remains the same, and the RN black hole with $N$ Planckons has the same entropy as the electrically neutral hole, $S_{\rm RNBH}(N)=N(N-1)/2$. This is supported by the adiabatic process of transformation from the RN to Schwarzschild black hole by varying the fine structure constant. The adiabaticity is violated in the extreme limit, when the gravitational interaction between two Planckons is compensated by the repulsion between their electric charges, and the RN black hole loses stability. The entropy of a white hole formed by the same $N$ Planckons has negative entropy, $S_{\rm WH}(N)=-N(N-1)/2$. 
\end{abstract}
\pacs{
}

\maketitle

\tableofcontents

 \section{Introduction}

In references \cite{Volovik2025a,Volovik2025d}  we considered the black hole thermodynamics using the non-extensive statistics. The entropy of the black hole obeys the composition rule
\begin{equation}
\sqrt{S_{\rm BH}(M_1 +M_2)}= \sqrt{S_{\rm BH}(M_1)} +\sqrt{S_{\rm BH}(M_2)}\,,
\label{BlackHoles}
\end{equation}
 which coincides with the composition rule in the non-extensive Tsallis-Cirto $\delta=2$ statistics.\cite{TsallisCirto2013,Tsallis2020}
Eq.(\ref{BlackHoles}) expresses the entropy of the black hole with mass $M_1+M_2=M$, which is obtained by the merging of two black holes with masses  $M_1$ and $M_2$. In the process of merging the entropy increases. The reversed process of splitting of the black hole with mass $M$ into two smaller black holes with $M_1+M_2=M$ has been calculated using the quantum tunneling approach developed 
by Parikh and Wilczek.\cite{Wilczek2000}  These calculations showed that the rate of this process can be expressed in terms of the difference in entropy before and after splitting, $w\propto \exp(-\Delta S)$, which confirms the composition rule in the equation (\ref{BlackHoles}). Then, following the Landau-Lifshitz book "Statistical Physics"\cite{Landau_Lifshitz}, one can assume that the splitting of a black hole in the process of macroscopic quantum tunneling is equivalent to a thermodynamic fluctuation in which the entropy decreases after splitting and then is restored by merging.

Here we consider a toy model, which gives rise to quantization of the generalized black hole entropy. The black hole is considered as an ensemble of the Planck mass objects -- Planckons.\cite{Markov1967,Hawking1971,Aharonov1987,Rovelli2024,YenChinOng2025,Trivedi2025}
If Planckon has the reduced Planck mass, $m_{\rm P}=1/\sqrt{8\pi G}$, the black hole with mass $M=Nm_{\rm P}$ consisting of $N$ Planckons has the integer valued entropy $S_{\rm BH}(N)=N(N-1)/2$. This leads to the extension of the composition rules, in which the processes of merging and splitting of black holes are expressed in terms of the integer numbers. For example, the probability of splitting of a black hole into two smaller parts, $N=N_1+N_2$, is given by the exponent $\exp(-\Delta S)=e^{-N_1N_2}$. For $N_1=1$ and $N_2=N-1$ this gives the rate of emission of single Planckons. This rate automatically includes the effect of back reaction in the process of radiation, which was first considered by Parikh and Wilczek.\cite{Wilczek2000}

We also consider the extension of the generalized entropy to white holes and to the Reissner-Nordstr\"om (RN) black holes. We find that the statistical ensemble of charged $N$ Planckons forming the RN black hole has the same entropy as the electrically neutral black hole, $S_{\rm RNBH}(N)=S_{\rm BH}(N)=N(N-1)/2$. This is valid up to the extreme limit, after which the RN black hole loses stability.
For a white hole, the anti-symmetry with respect to time reversal gives negative entropy, $S_{\rm WH}(N)=-S_{\rm BH}(N)=-N(N-1)/2$.
 
\section{Schwarzschild black hole}  
\label{SchwarzschildSec}
 
\subsection{Horizon, irreversibility and thermodynamics}  
\label{HorizonSec}
   
 We use the Painleve-Gullstrand coordinates,\cite{Painleve,Gullstrand} where  the metric has the following form:
\begin{equation}
ds^2= - dt^2(1-{\bf v}^2) - 2dt\, d{\bf r}\cdot {\bf v} + d{\bf r}^2 \,.
\label{PGmetric}
\end{equation}
Here the vector ${\bf v}$ is the shift function in the Arnowitt-Deser-Misner formalism.\cite{ADM2008}, while in the condensed matter analogs this vector corresponds to the "velocity of the quantum vacuum". 
 For the Schwarzschild black and white holes the shift function is:
\begin{equation}
{\bf v} ({\bf r})=\mp \hat{\bf r} \sqrt{\frac{2MG}{r}}\,.
\label{velocity}
\end{equation}
The minus sign in the equation (\ref{velocity}) corresponds to the black hole configuration (the velocity ${\bf v} $ is directed toward the horizon), and the plus sign describes a white hole (the "vacuum" flows away from the horizon). In both cases the event horizon is at $r=2MG$. 

It is the event horizon which is responsible for irreversibility and thus for the thermodynamic behavior of the black and white holes.\cite{Witten2025} The external particle is absorbed by the black hole with probability 1, while the reversed process -- the emission from the black hole  -- is suppressed by the Hawking exponent describing the rate of the quantum tunneling from the black hole. The detailed balance gives rise to the Hawking temperature of the black hole and to the Hawking temperature with minus sign for the white hole. 

On the other hand the processes of merging and splitting of the black holes by macroscopic quantum tunneling determine the entropy of these objects. They obey the Tsallis-Cirto statistics, where the entropy is proportional to the square of the logarithm of probability:
\begin{equation}
S_{\delta =2}= \sum_i p_i \left(\ln\frac{1}{p_i} \right)^2\,.
\label{PlusEntropy}
\end{equation}

\subsection{Entropy in terms of Planck-scale elements}  

This peculiar type of the Tsallis-Cirto statistics (i.e. with $\delta=2$ in Eq.(\ref{PlusEntropy})) also suggests that the black hole can be represented as ensemble of $N$ "black hole atoms" -- the smallest collapsed objects introduced by Hawking,\cite{Hawking1971} 
or "Planckons",\cite{Aharonov1987,YenChinOng2025} see also Markov's maximons.\cite{Markov1967} Planckon  is the object with the Planck scale mass. The proper choice of the Planckon mass, which naturally fits the  the Tsallis-Cirto statistics, is the reduced Planck mass, $m_{\rm P}^2=1/8\pi G$, so that the mass of the black hole is \begin{equation}
M=N m_{\rm P} \,\,,\,\,m_{\rm P}=1/\sqrt{8\pi G}\,,
\label{BWHolesN2}
\end{equation}

This allows the Tsallis-Cirto entropy to be expressed in terms of an integer number of Planckons:\begin{equation}
S_{\rm BH}(N)=\begin{pmatrix}N
\\ 2\end{pmatrix} =\frac{N! }{2! (N-2)!}=\frac{N(N-1)}{2}  \,\,, \,\, N=\frac{M}{m_{\rm P}}\,.
\label{BWHolesN2}
\end{equation}
This entropy coincides with the Bekenstein-Hawking entropy in the limit of large $N$. On the other hand, it also describes the entropy of an ensemble with finite $N$.

 In Eq.(\ref{BWHolesN2}), the entropy of the ensemble arises from the correlations between Plancons. It can be interpreted as the number $\cal N$ of ways to select two Planckons from an ensemble of $N$ Planckons:
\begin{equation}
S_{\rm BH}(N) ={\cal N}=\frac{N(N-1)}{2}  \,.
\label{BWcalN}
\end{equation} 
In this interpretation, $N$ Planckons interact gravitationally with each other, and their correlated (or entangled) pairs provide ${\cal N}$ thermodynamic degrees of freedom for the black hole. A single Planckon has no partner to form a pair with, and so has zero entropy:
\begin{equation}
 S(N=1)=S({\cal N}=0)=0  \,.
\label{ZeroEntropy}
\end{equation} 
The zero entropy of an individual Planckon distinguishes these toy objects from physical Planck-scale black holes.

That it is the pairs, and not the Planckons themselves, that are the black hole's thermodynamic degrees of freedom can be seen from the equipartition law, which holds for large ${\cal N}$:
\begin{equation}
M=2{\cal N}T_{\rm H} \,\,,\,\, {\cal N}\gg 1 \,.
\label{equipartition}
\end{equation}
Here $T_{\rm H}$ is the Hawking temperature:
\begin{equation}
T_{\rm H} =\frac{1}{8\pi GM}=\frac{m_{\rm P}}{N}\,.
\label{HawkingT}
\end{equation}

\subsection{Splitting of black hole into smaller parts and back reaction}  

The discrete version of the Tsallis-Cirto $\delta=2$ entropy determines the processes of merging and splitting of ensembles of Planckons.
For example, the probability of a black hole splitting into two smaller parts, $N=N_1+N_2$, in the process of macroscopic quantum tunneling is determined by the decrease in entropy after the split:
\begin{equation}
w(N\rightarrow N_1+N_2)= \exp[S_{\rm BH}(N_1)+S_{\rm BH}(N_2) - S_{\rm BH}(N_1+N_2)] =\exp(-N_1N_2)\,.
\label{splitting}
\end{equation}
In particular, the rate of emission of one Plankcon by a black hole is:
\begin{equation}
w(N\rightarrow (N-1)+1)=e^{-(N-1)}  \,.
\label{radiation}
\end{equation}
It can be written in the form corresponding to thermal radiation of Planckons with Hawking temperature:
\begin{equation}
w=e^{-(N-1)} =\exp\left(-\frac{m_{\rm P}}{T_{\rm H}}\left(1-\frac{m_{\rm P}}{M}\right)\right) \,.
\label{Hawkingradiation}
\end{equation}
The correction to the Hawking radiation rate corresponds to the effect of back reaction in the process of radiation. This effect was first considered by Parikh and Wilczek,\cite{Wilczek2000}  the  extension of their theory to the processes of splitting of the black holes see in Section II.F of Ref. \cite{Volovik2022}.

Similarly, we can consider the processes of division of a black hole into more than two parts. The probability of division into four smaller parts contains $4\times (4-1)/2=6$ exponents:
\begin{equation}
w(N\rightarrow N_1+N_2+N_3+N_4)=\exp(-N_1N_2-N_1N_3 -N_1N_4-N_2N_3 -N_2N_4-N_3N_4)\,.
\label{splitting4}
\end{equation}
Then the splitting of a black hole into $k$ black holes contains $k/(k-1)/2$ exponents, and
finally, the probability of splitting of the black hole into its $N$ Planckons  contains ${\cal N}=N(N-1)/2$ exponents: 
\begin{equation}
w(N\rightarrow \underbrace{1+1+...+1}_N \,)=e^{-\cal N}=\exp\left(-\frac{N(N-1)}{2}\right)=\exp(-S_{\rm BH}(N))\,. 
\label{splittingN}
\end{equation}


According to Eq.(\ref{radiation}), this process can be considered as the sequence of radiation of Planckons one by one until the complete destruction of the black hole:
\begin{equation}
w(N\rightarrow \underbrace{1+1+...+1}_N \,)=\Pi_{k=1}^{N}e^{-(N-k)}=\exp\left(-\frac{N(N-1)}{2}\right)=\exp(-S_{\rm BH}(N))\,. 
\label{radiationN}
\end{equation}
In this process the entropy decreases from $S_{\rm BH}(N)=N(N-1)/2$ to $S=0$. 

Equations (\ref{splittingN}) and (\ref{radiationN}) give the following physical meaning of the entropy of a black hole, regardless of the model. The entropy of a black hole determines the probability of the complete annihilation of the black hole.

\section{White holes} 

\subsection{Probability of formation of white hole and negative entropy}  

The white hole has extremely fine-tuned nature as was discussed by Witten.\cite{Witten2025,Eardley1974} The formation of the white hole from $N$ Planckons is extremely unlikely. The white hole metric is obtained from the metric of the black hole by the time reversal, at which the shift vector ${\bf v}$ of PG coordinates changes sign, $T{\bf v}=-{\bf v}$. Due to this time anti-symmetry between the black and white holes, the probability of formation of the white hole by combining $N$ Planckons is the same as probability of black hole splitting to $N$ Planckons in Eqs. (\ref{splittingN}) and (\ref{radiationN}): 
 \begin{equation}
w(\,\underbrace{1+1+...+1}_N \,\rightarrow  N)=\exp\left(-\frac{N(N-1)}{2}\right)=e^{-\cal N}\,.
\label{inversN}
\end{equation}
This means that formation of the white hole by merging of Planckons represents the rare thermodynamic fluctuation, at which the entropy decreases from $S=0$ to the maximum negative value, which corresponds to the entropy of a black hole with a minus sign:
\begin{equation}
S_{\rm WH}(N)=-\frac{N(N-1)}{2}=-{\cal N}=-S_{\rm BH}(N)   \,\,, \,\, N=\frac{M}{m_{\rm P}}\,.
\label{WHHolesN2}
\end{equation}

Eq. (\ref{WHHolesN2}) is supported by the direct calculations of the transition rate from the black hole to the white hole by macroscopic quantum tunneling,\cite{Volovik2022} which demonstrate that the entropy of the white hole with mass $M$ is with minus sign the entropy of the black hole with the same mass:
\begin{equation}
 S_{\rm WH}(M)=-S_{\rm BH}(M) =-4\pi GM^2
\,.
\label{WHBH}
\end{equation}

 \subsection{White hole and modified Bekenstein entropy bound}  

Eq. (\ref{WHHolesN2}) suggests the analog of Bekenstein entropy bound\cite{Bekenstein1981,Bekenstein2004} and holographic entropy bound\cite{Hooft1993,Susskind1995}, which now includes the negative entropy of white holes. The entropy of an ensemble of $N$ Planckons is bounded by maximum and minimum values that correspond to the entropies of black and white holes:
\begin{eqnarray}
-\frac{N(N-1)}{2} \leq S(N) \leq \frac{N(N-1)}{2}\,.
\label{Bound}
\end{eqnarray}
The entropy of a white hole is the negative entropy of the least probable exceptional state. 
The white hole is unstable to the radiation of Planckons, and in the process of evolution all Planckons leave the white hole. Then these Planckons are attracted to each other and eventually form a black hole. We assume here that the system is finite, so that Planckons cannot escape to infinity after Hawking radiation, otherwise the black hole will finally decay. 

In this scenario, as in other scenarios of evolution, the white hole relaxes into a black hole with increasing entropy. This demonstrates that due to the negative entropy of the white hole, the second law of thermodynamics is not violated. On the other hand, the inverse process, i.e. the process of formation of white hole, corresponds to quantum fluctuation with probability $w \sim \exp\left(-N(N-1)\right)$.

 \subsection{Tsallis-Cirto statistics with negative entropy and second law of thermodynamics}  
 
 In all the processes of splitting of the black hole, the entropy decreases, as it does in rare thermodynamic fluctuations.\cite{Landau_Lifshitz} The process of macroscopic quantum tunneling, in which a black hole transforms into a white hole, can also be viewed as a thermodynamic fluctuation in which the entropy decreases. Thermodynamic fluctuations are rare events, they do not violate the second law of thermodynamics since this negative jump of the entropy after splitting is compensated by the further processes of absorption of the radiated objects which restore the entropy of the black hole. 

On the other hand, the natural process of splitting of the white hole into smaller parts is always accompanied by increase of entropy. The increase of entropy does not require fluctuations in the form of macroscopic quantum tunneling. That is why the white hole also does no violate the second law of thermodynamics.

So, the black and white holes are thermodynamic objects of special type. Due to the presence of horizon, they obey the special type of the thermodynamic statistics --  the extended Tsallis-Cirto $\delta=2$ statistics with 
\begin{equation}
S^\pm_{\delta =2}=\pm \sum_i p_i \left(\ln\frac{1}{p_i} \right)^2\,.
\label{PlusMinusEntropy}
\end{equation}
 This differs from both the usual von Neumann statistics, and also from the original $\delta=2$ Tsallis-Cirto statistics in Eq.(\ref{PlusEntropy})) by allowing a negative value for the entropy of the white-hole horizon.\cite{Volovik2025d} States with negative thermodynamic entropy are forbidden in ordinary thermodynamics of systems without horizons, but are natural for thermodynamics of systems with horizons.

 \section{Reissner-Nordstr\"om black hole} 
 \label{RNsection} 
 
 \subsection{Negative entropy of inner horizon and composition law}

 In the extended Thallis-Cirto statistics in Eq.(\ref{PlusMinusEntropy}), the entropy of Reissner-Nordstr\"om (RN) black hole can be considered as the composition of the entropies of the outer and inner horizons:
 \begin{equation}
S_{\rm RN}=\left(\sqrt{S_{\rm RN}(r_+)}+\sqrt{|S_{\rm RN}(r_-)|} \right)^2 =4\pi GM^2 \,.
\label{compositionRN}
\end{equation}
 Here entropy of the outer horizon is
\begin{equation}
S_{\rm RN}(r_+)= \pi r_+^2/G=\pi G \left(M+ \sqrt{M^2 -\alpha Q^2} \right)^2\,,
\label{outerRN}
\end{equation}
where $\alpha$ is the fine structure constant, and $Q$ is the integer valued electric charge with $Q=-1$ for electron, and the entropy of the inner horizon is negative:
 \begin{equation}
S_{\rm RN}(r_-)= -\pi r_-^2/G=-\pi G \left(M- \sqrt{M^2 -\alpha Q^2} \right)^2\,.
\label{innerRN}
\end{equation}

The minus sign for the entropy of the inner horizon in Eq. (\ref{innerRN}) is due to the fact that the inner horizon is equivalent to the white hole horizon and therefore its entropy is negative. The negative entropy of the inner horizon also agrees with the negative temperature of the inner horizon, $T_-= - (r_+ -r_-)/4\pi r_-^2$, see Eq.(53) in Ref. \cite{Volovik2022}.

The composition rule for the Reissner-Nordstr\"om black hole shows that the total entropy of an RN black hole is completely determined by its mass $M$. This is consistent with the adiabatic transformation of the RN black hole into a Schwarzschild black hole, where the parameter $\alpha$ (the fine structure constant) adiabatically transforms to zero for fixed $M$ and $Q$. During this adiabatic transformation, the entropy does not change.

In the Planckon description, the RN black hole can be considered as ensemble of of Planckons, each with charge $q=Q/N$ (the charged Planckons have been also considered by Markov.\cite{Markov1967}).
Then Eq.(\ref{compositionRN}) demonstrates that the entropy of the RN ensemble does not depend on the Planckon charge. The entropy of RN black hole depends only on the number ${\cal N}$ of degrees of freedom, and thus on the number $N$ of Planckons. 
 \begin{equation}
S_{\rm RN} ={\cal N}=\frac{N(N-1)}{2}\,.
\label{NcompositionRN}
\end{equation}

 \subsection{Extremality and fine structure constant}
 
Equation (\ref{NcompositionRN}) is valid if the RN black hole is non-extremal. The transition to a super-critical black hole by continuously changing the fine structure constant $\alpha$ is non-adiabatic, as can be seen from the entropy $S_+$ and $S_-$ of outer and inner horizons:
\begin{equation}
S_\pm  =\pm  \frac{N(N-1)}{8}\left(1 \pm \sqrt{1-8\pi\alpha q^2}  \right)^2 \,\,, \,\, q=\frac{Q}{N} \,.
\label{RNentropues}
\end{equation}
The continuity is violated at the critical value, $\alpha_c= 1/(8\pi q^2)$. This discontinuity can be explained in terms of gravitational and Coulomb interactions between Planckons: 
\begin{equation}
U(r) = \frac{\alpha q^2}{r} - \frac{Gm_{\rm P}^2}{r}  \,.
\label{Interaction}
\end{equation}
At a critical value $\alpha=\alpha_c$ the interaction disappears, $U(r,\alpha_c)=0$. At larger $\alpha>\alpha_c$ the interaction between Planckons becomes repulsive, which leads to instability. The Planckons fly apart, emptying the black hole, and the entropy drops to zero.

At fixed $\alpha$ the transition to super-criticality is determined by the critical charge of Planckon:
\begin{equation}
q_c=\frac{1}{\sqrt{8\pi \alpha}} \,.
\label{CriticalCharge}
\end{equation}
For the current value of the fine structure constant, the critical charge is about two electron charges, $q_c\sim 2$.

 \subsection{Horizons, thermodynamics and Lifshitz transition}
 
So, the approach to extremality is not continuous,\cite{Carroll2009} and the entropy of the RN black hole drops from ${\cal N}=N(N-1)/2$ to zero when two horizons merge and annihilate. In the same manner the entropy of the RN white hole jumps from its negative value $-{\cal N}=-N(N-1)/2$ to zero when horizons annihilate. This shows that the negative entropy only holds in the presence of horizons.

At the critical point, at $\alpha_c=1/8\pi q^2$, adiabatic principle is violated.  The discontinuous transition between two regimes can be considered as a type of topological quantum phase transitions\cite{Volovik2007} in which horizons are formed, destroyed, annihilate, or change their topology.\cite{Jacobson1995,Smolin2003} Here the horizon represents the space-time analogue of the Fermi surface, which undergoes Lifshitz transitions.\cite{Lifshitz1960}

The event horizon is the source of irreversibility, which is similar to the statistical irreversibility.\cite{Witten2025} The emergence of horizons corresponds to the spontaneous breaking of time reversal symmetry, which leads to irreversibility. Without horizons, PG metrics with opposite shift vectors ${\bf v}$ describe the same state, since the two metrics can be transformed into each other by continuous coordinate transformations. This is why the metric becomes symmetric under time reversal and thus there is no irreversibility. It is the event horizons that provide positive temperature and entropy for a black hole and negative temperature and entropy for a white hole. This can be seen from the specific detailed balance in the presence of horizon:
\begin{eqnarray}
P_{\rm emission}=P_{\rm absorption}\exp(-\omega/T) \,,
\label{Balance}
\end{eqnarray}
where for simplicity we consider the limit $\omega\gg T$. 

Here by absorption we mean the absorption of external matter by the black hole, which is equivalent to conventional spontanous emission from an excited state. In the presence of a black hole horizon, matter is inevitably absorbed with probability $P_{\rm absorption}=1$, and emission occurs only through quantum tunneling from the black hole with probability
$P_{\rm emission}=\exp(-\omega/T_{\rm Hawking})$. Then the detailed balance (\ref{Balance}) gives the Hawking temperature for black hole, $T=T_{\rm Hawking}=1/8\pi M$. For the white hole horizon the emission is inevitable, $P_{\rm emission}=1$, while now absorption is provided by quantum tunneling to white hole, $P_{\rm absorption}=\exp(-\omega/T_{\rm Hawking})$. This corresponds to negative temperature $T=-T_{\rm Hawking}=-1/8\pi M$.

In the case of a RN black hole we have $P_{\rm absorption}=1$ and $P_{\rm emission}=\exp(\omega/|T_-|)\exp(-\omega/T_+)$. Then the detailed balance (\ref{Balance}) gives $1/T=1/T_+  - 1/|T_-|=4\pi(r_+ + r_-) =8\pi M$. This again supports the conclusion that the thermodynamics of the RN black hole coincides with the thermodynamics of the electrically neutral black hole with the same mass.

 \section{de Sitter thermodynamics in Planckon model and quantization of $\Lambda$} 
 \label{dSsection} 
 
  \subsection{Thermodynamics of cosmological horizon}
  
In the Painleve-Gullstrand coordinates, the corresponding metric has the shift vector:
\begin{equation}
{\bf v}=H{\bf r} \,,
\label{PGmetric}
\end{equation}
where $H$ is the Hubble parameter.

The processes associated with the cosmological horizon at $R=1/H$ also obey the balance equation (\ref{Balance}). The corresponding temperature governing the processes of radiation from the horizon is the Gibbons-Hawking temperature $T_{\rm GH}=H/2\pi$. However, the real local temperature of the de Sitter state which describes the de Sitter thermodynamics, is not related to horizon and is twice as large,
$T=H/\pi=2T_{\rm GH}$.\cite{Volovik2025b} This local temperature determines the local entropy density $s$ of the de Sitter state. The total entropy of the Hubble volume $V_H$ (the volume inside the cosmological horizon) nevertheless coincides with the Gibbons-Hawking entropy $S_{\rm GH}$ of the cosmological horizon:
\begin{eqnarray}
S_H=sV_H=\frac{A}{4G}=S_{\rm GH}\,,
\label{HubbleVolume}
\end{eqnarray}
 where $A$ is the area of the cosmological horizon. 
 Thus, although the local temperature does not coincides with the Gibbons-Hawking temperature, this does not violate the holographic bulk-surface correspondence.
 The factor two between the local temperature $T$ and the temperature $T_{\rm GH}$ associated with the cosmological horizon is a consequence of a special symmetry of the de Sitter state and is supported by topological arguments.\cite{Kusmartsev2025,Volovik2025f}
 
Another connection is provided by the Planckon toy model. According to this model, the energy and entropy of the Hubble volume are quantized in the same way as in the case of a black hole: 
\begin{eqnarray}
E_H=N m_{\rm P} \,\,,\,\, S_H ={\cal N}=\frac{N(N-1)}{2} \,.
\label{Hubble}
\end{eqnarray}
Here $N$ is the number of Planckons in the Hubble volume.

For large $N\gg 1$ the entropy of the Hubble volume approaches the Gibbons-Hawking entropy:
\begin{eqnarray}
S_H(N\gg 1) \rightarrow \frac{1}{2}N^2= \frac{\pi}{GH^2}=\frac{A}{4G}\,.
\label{HubbleGibbons}
\end{eqnarray}

 \subsection{Quantization of cosmological constant}

While the quantized entropy of the Hubble volume is a rather abstract quantity, the Planckon model suggests also the quantization of the physical observable -- the cosmological constant $\Lambda$. This is the energy density $\rho_{\rm vac}$ of the de Sitter state, which is a local quantity. From Planckon model one obtains:
\begin{eqnarray}
\Lambda\equiv \rho_{\rm vac}= \frac{3}{8 G^2{\cal N}}=\frac{24 \pi^2}{l_{\rm P}^4 {\cal N}}\,.
\label{Lambda}
\end{eqnarray}
Here $l_{\rm P}=1/m_{\rm P}$ is the reduced Planck length. A similar quantization of $\Lambda$ is obtained in the gravitational-topological sector.\cite{Wieland2011,Alexander2025}
Equation (23) in Ref.\cite{Wieland2011} implies for the reduced cosmological constant the values
$\Lambda=\frac{12\pi}{\beta l_{\rm P}^2 n}$, where $\beta$ is the Barbero–Immirzi parameter and $n$ is an integer. With $\beta=\frac{1}{2\pi}$ and $n=\cal N$, this agrees with Eq.(\ref{Lambda}).

Let us also mention the quantum theory of gravity, in which tetrads are formed as bilinear combinations of fermion fields and have the dimension of the inverse length.\cite{Diakonov2012} As a result, the cosmological constant $\Lambda$ is dimensionless, since $l_{\rm P}^4$ is absorbed by the 4-from tetrad determinant.  For dimensionless physical quantities, quantization is quite natural. In this theory of gravity, the covariant form of the scalar area is also dimensionless, which makes quantization of the horizon area and thus of the horizon entropy also rather natural.\cite{Volovik2021}

 \subsection{Negative entropy of contracting de Sitter}
 
The time reversal transforms the expanding de Sitter with the positive Hubble parameter, $H>0$, to contracting de Sitter with $H<0$. Correspondingly, due to anti-symmetry of the Hubble parameter, $TH=-H$, the entropy of the volume inside the cosmological horizon (the Hubble volume) is negative in the contracting Universe,\cite{Volovik2025b}
\begin{eqnarray}
S(H<0) =-S(H>0)= -\frac{A}{4G}\,.
\label{HubbleVolumeNeg}
\end{eqnarray}

 Transition between expanding and contracting states of de Sitter Universe may occur via Minkowski state. In this scenario, the local entropy density $s$ changes sign by crossing the zero value, which is natural for the empty Minkowski vacuum with temperature $T=0$.
But at the transition point, i.e. in  Minkowski vacuum, the horizon is at $r=\infty$. This means that in this transition, the total entropy $S(H)$ of the Hubble volume $V_H$ changes sign by crossing an infinite value in the Minkowski vacuum. This demonstrates that spatial infinity in the Minkowski vacuum can play a role similar to that of the cosmological event horizon. Escape to infinity looks similar to escape through the crossing of the cosmological horizon. 

The Planckon model also suggests that Minkowski vacuum is not empty. While the density of Planckons $N/V_H=3H^2m_{\rm P}$ approaches zero in the Minkowski limit, their total number in the Hubble volume, $N=\frac{4\pi m_{\rm P}}{H}$, approaches infinity at $H\rightarrow 0$ (see also Ref.\cite{Carr2025} on the decay of de Sitter space into real physical Planckons). This is the consequence of the holographic bulk-surface correspondence.
The holography and possible escape to space infinity raise the problem of decoherence in the Minkowski vacuum,\cite{Danielson2023,Danielson2025} and probably are important for the problem of the collapse of the wave function.\cite{Bassi2013,Bose2025} 

 \section{Conclusion}
 
 Thermodynamics of black and white holes can be considered in the toy model of the ensemble of the Planck mass objects, Planckons. The Tsallis-Cirto non-extensive statistics with $\delta=2$, which describes the processes of splitting and merging of black holes, suggests the statistical properties of this ensemble. The entropy of black hole with $N$ Planckons is provided by the $N(N-1)/2$ degrees of freedom -- the correlations between the gravitationally attracted Planckons. If one chooses the mass of Planckon as the reduced Planck mass, $m_{\rm P}=1/\sqrt{8\pi G}$, and thus the mass of the black hole is $M=Nm_{\rm P}$, then the entropy of this black hole is $S_{\rm BH}(N)=N(N-1)/2$. 
 
 This describes the processes of splitting and merging of black holes in integer numbers. For example, the probability of splitting of the black hole with $N$ Planckons to the two black holes $N_1+N_2=N$ is
  $w\propto e^{-N_1N_2}$. For $N_1=1$ and $N_2=N-1$ this gives the probability of thermal radiation of Planckon with Hawking temperature $T_{\rm H}=1/8\pi GM$ supplemented by the back-reaction effect in the radiation process, which first considered by Parikh and Wilczek. The probability of the decay of a black into its $N$ Planckons is $w\propto e^{-N(N-1)/2}$, which relates the entropy of a black hole to the probability of its complete annihilation.
 
The entropy of a white hole is negative as follows from the rate of macroscopic quantum tunneling from a black hole to a white hole. In terms of Planckons, the entropy of a white hole is the negative of the entropy of a black hole with the same number of Planckons, $S_{\rm WH}(N)=-N(N-1)/2= -S_{\rm BH}(N)$.  This reflects the anti-symmetry with respect to time reversal, at which the shift vector in the Arnowitt-Deser-Misner formalism changes sign. This anti-symmetry allows one to extend the Tsallis-Chirto entropy by adding a minus sign to the Tsallis-Chirto formula applied to white hole.  
 
The Planckon model has been extended to a Reissner-Nordstr\"om (RN) black hole, which consists of the electrically charged Planckons. The entropy of a Reissner-Nordstr\"om black hole is obtained as a non-extensive composition of the positive entropy of the outer horizon and the negative entropy of the inner horizon.  Despite the charge, the statistical ensemble of Planckons remains the same, and the RN black hole with $N$ Planckons has the same entropy as the electrically neutral hole, $S_{\rm RN}(N)=N(N-1)/2$. This is confirmed by the adiabatic process of transition from RN to a Schwarzschild black hole by continuous changing the fine structure constant. Adiabaticity is broken in the extreme limit, when the gravitational interaction between two Planckons is compensated by the repulsion between their electric charges, and the RN black hole loses stability and disintegrates into Planckons.
 
The Planckon model, when extended to the de Sitter state of the quantum vacuum, implies quantization of the cosmological constant $\Lambda$. This connects the thermodynamics of the cosmological horizon with the local thermodynamics of the de Sitter state, as well as with the topological sector of quantum gravity.
 
According to Unruh,\cite{Unruh2024} the quantum mechanics of black holes is still a far from finished topic.
Also according to Jacobson,\cite{Jacobson2024} the connection between the de Sitter horizon entropy and quantum field theory still stands on shaky ground. The Planckon toy model will likely give us a deeper understanding of the origins of thermodynamics that arise in black holes, as well as in the quantum vacuum.

\end{document}